\documentstyle[bo99,epsfig]{article}
\def\ls{{_<\atop^{\sim}}}

\def\gs{{_>\atop^{\sim}}}

\def\cgs{ ${\rm erg~cm}^{-2}~{\rm s}^{-1}$ } 

\title{The BeppoSAX HELLAS survey: on the nature of faint hard X-ray selected
sources}

\author{F.~Fiore$^{1,2,3}$, L.A.~Antonelli$^2$, P.~Ciliegi$^4$, 
A.~Comastri$^4$, P.~Giommi$^1$, \\F.~La~Franca$^5$, R.~Maiolino$^6$, 
G.~Matt$^5$, S.~Molendi$^7$,  G.C.~Perola$^5$, C.~Vignali$^4$}
\affil{1) BeppoSAX Science Data Center, Via Corcolle 19, I--00131 Roma, 
Italy; \\
2) Osservatorio Astronomico di Roma, Via Frascati 33,
I-00044 Monteporzio, Italy;\\
3) Harvard-Smithsonian Center of Astrophysics, 60 Garden Street, 
Cambridge MA 02138 USA;\\
4) Osservatorio Astronomico di Bologna, via Ranzani 1, I40127
Bologna, Italy;\\
5) Dipartimento di Fisica, Universit\`a degli Studi ``Roma Tre",
Via della Vasca Navale 84, I--00146 Roma, Italy;\\
6) Osservatorio Astrofisico di Arcetri, p. E. Fermi 5, Firenze, Italy;\\
7) IFCTR/CNR, via Bassini 15, Milano, I20133, Italy}

\begin{document}

\maketitle

\begin{abstract}
The BeppoSAX 4.5-10 keV High Energy Large Area Survey has covered
about 80 deg$^2$ of sky down to a flux of
$F_{5-10keV}\sim5\times10^{-14}$ \cgs.  Optical spectroscopic
identification of $\ls$half of the sources in the sample (62) shows
that many ($\approx50\%$) are highly obscured AGN, in line with the
predictions of AGN synthesis models for the hard X-ray background
(XRB, see e.g. Comastri et al. 1995). The X-ray data, complemented by
optical, near-IR and radio follow-up, indicate that the majority of
these AGN are ``intermediate'' objects, i.e. type 1.8-1.9 AGN,`red'
quasars, and even a few broad line, blue continuum quasar, obscured in
X-rays by columns of the order of $10^{22.5-23.5}$ cm$^{-2}$, but
showing a wide dispersion in optical extinction. The optical and
near-IR photometry of the obscured objects are dominated by galaxy
starlight, indicating that a sizeable fraction of the accretion power
in the Universe may actually have been missed in optical color
surveys. This also implies that multicolor photometry techniques may
be efficiently used to assess the redshift of the hard X-ray selected
sources.
\end{abstract}

\section{Introduction}

Hard X-ray observations are the most efficient way of tracing emission
due to accretion mechanisms, such in Active Galactic Nuclei (AGN).
Hard X-ray selection is less affected by strong biases present at
other wavelengths. For example, a column of a few times $10^{22}$
cm$^{-2}$ has negligible effect in the 5-10 keV band, while it reduces
by $\sim100$ times nuclear emission below 2 keV. Soft X-ray surveys
(e.g. Hasinger et al. 1998, Schmidt et al. 1998) are also often
contaminated by non nuclear components, like emission from binaries
and/or from optically thin plasmas in star-formation regions
surrounding the nucleus.  Optical and UV color selection is biased
against objects with even modest extinction or an intrinsically `red'
emission spectrum (see e.g.  Vignali et al. 2000).  Sensitive hard
X-ray surveys are therefore powerful tools to select large samples of
AGN less biased against absorption and extinction.  Our approach
consists in taking advantage of the large field of view and good
sensitivity of the BeppoSAX MECS instrument (Boella et al. 1997a,b) to
survey tens to hundreds of square degrees at fluxes
$\gs5-10\times10^{-14}$ \cgs (Fiore et al. 2000a), and using higher
sensitivity XMM-Newton and Chandra observations to extend the survey
down to $\sim10^{-14}$ \cgs on several deg$^2$ (at this flux the
majority of the hard XRB is resolved in sources). The results on the
optical identification of a sample of faint Chandra sources discovered
over the first 0.14 deg$^2$ have been published by Fiore et
al. (2000b).  This approach is complementary to deep pencil beam
surveys ($\sim0.1$ deg$^2$, see e.g. Mushotzky et al. 2000,
Hornschemeier et al. 2000), and we cover a different portion of the
redshift--luminosity plane.  Our purpose is to study cosmic source
populations at fluxes where a reasonably large fraction of the hard
XRB is resolved (20-30\% at the BeppoSAX flux limit), but where the
X-ray flux is high enough to provide X-ray spectral information in
higher sensitivity follow-up observations. This would allow the
determination of the distribution of absorbing columns in the sources
making the hard XRB, providing strong constraints on AGN synthesis
models for the XRB. Furthermore, large area surveys allow the search
for previously `rare' AGN, like `red' quasars or other minority AGN
populations (Kim \& Elvis 1999) and quantify their fractional
contribution to the AGN family.  Finally, at our X-ray flux limits the
optical counterparts are bright enough to allow relatively high
quality optical spectroscopy, useful to investigate the physics of the
sources.

\section{The HELLAS survey}

The High Energy Large Area Survey (HELLAS) has been performed in the
4.5-10 keV band because: a) this is the band closest to the maximum of
the XRB energy density which is reachable with the current imaging
X-ray telescopes, and b) the BeppoSAX MECS Point Spread Function (PSF)
greatly improves with energy, providing a 95 \% error radius of 1$'$
in the hard band (Fiore et al. 2000a), allowing optical identification
of the X-ray sources. About 80 deg$^2$ of sky have been surveyed so
far using 142 BeppoSAX MECS fielda at $|b|>20$ deg.  (Fields centered
on bright extended sources and bright Galactic sources were excluded
from the survey, as well as fields close to LMC, SMC and M33.)

A robust detection algorithm has been used in coadded
MECS1+MECS2+MECS3 (or MECS2+MECS3 for observations performed after the
failure of the MECS1 unit) 4.5-10 keV images. The method consists in
first convolving the X-ray image with a wavelet function, to smooth
the image and increase contrast, and then in running a standard
slide-cell detection method on the smoothed image, to locate count
excesses above the local background.  The statistics of each candidate
detection is then accurately studied and the final net counts are
estimated from the original (un-smoothed) image to preserve Poisson
statistics.  The background is calculated using source-free boxes near
the source region.  The detection has been run several time for each
field, changing the size of the wavelet function, to a) take into
account the variation of the MECS PSF with the offaxis angle, and b)
to detect efficiently sources with variable extension.  The quality of
the detection has always been checked interactively.  We used a
probability threshold of 99.94 \% (about 3.5 $\sigma$).  Sources
detected in regions of radius 4, 6 or 8 arcmin around targets
(depending on the target brightness) have been excluded from the
sample, which includes a total of 147 sources down to a 5-10 keV flux
of $\sim5\times10^{-14}$ \cgs.  Count rates were converted to fluxes
using a fixed conversion factor equal to 7.8$\times10^{-11}$ \cgs
(5-10 keV flux) per one ``3 MECS count'' (4.5-10 keV). This factor is
appropriate for a power law spectrum with $\alpha_E=0.6$, but due to
the narrow band it is not strongly sensitive to the spectral shape:
for $\alpha_E=0.4$ and 0.8 it is 8.1 and 7.6 $\times10^{-11}$ \cgs
respectively.  The skycoverage varies from $\sim1$ deg$^2$ at
$5\times10^{-14}$ \cgs to 6.6, 50 and 84 deg$^2$ at fluxes of
$10^{-13}, 3\times10^{-13}$ and $>10^{-12}$ \cgs respectively. After
correcting for this skycoverage we find 16.9$\pm$3.0(6.4) sources
deg$^{-2}$ at $F_{5-10 keV}=4.8\times10^{-14}$ \cgs
(Fiore et al. 2000a, Comastri et al. 2000). First quoted
errors are the $1\sigma$ statistical confidence interval. Errors in
brackets include systematic uncertainties, due to the lack of
knowledge of the real spectrum of the faint sources.  This logN-logS
corresponds to a resolved fraction of the 5-10 XRB equal to 20-30 \%,
depending on the XRB normalization (Comastri this meeting, Vecchi et
al. 1999).  The observed number counts are consistent, within the
errors, with the extrapolations from the BeppoSAX (Giommi et al 2000)
and ASCA (Cagnoni et al. 1998, Ueda et al. 1999, Della Ceca et al.
1999) 2-10 keV number counts, assuming an average power law spectrum
with $\alpha_E\sim0.6$.

\section{Hardness ratio analysis}

\begin{figure}[ht]
\centerline{
\psfig{file=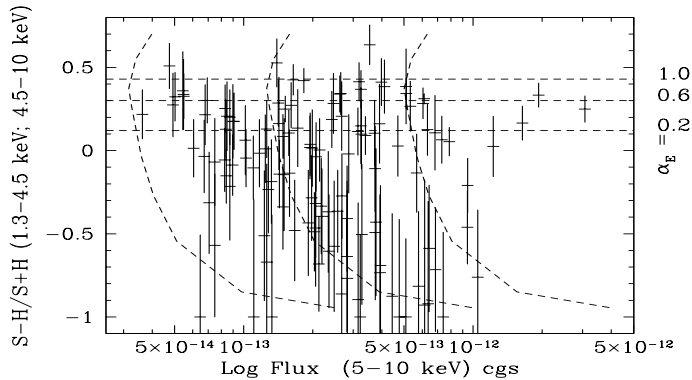, width=6cm, height=10cm, angle=-90}
}
\caption{
The softness ratio (S-H)/(S+H) versus the 
5-10 keV flux for the 126 sources in the HELLAS sample not
covered by the strongback support of the MECS window.
Dashed lines mark loci of equal 4.5-10 keV count rate}
\label{hrtfx}
\end{figure}

To study the spectral variety of the HELLAS sources we have calculated
for each source the softness ratio (S-H)/(S+H), S=1.3-4.5 keV,
H=4.5-10 keV.  Figure 1 plots (S-H)/(S+H) as a function of source
total 5-10 keV flux. Sources under or close to the berillium
strongback supporting the MECS window have been excluded from this
analysis, because their observed softness may be systematically lower
than real. The number of the remaining sources in the sample is 126.
Many of the HELLAS sources have a low (S-H)/(S+H), indicating a hard
spectrum.  Assuming $\alpha_E=0.6$, we find that 36(5) of the 126
sources have 1.3-4.5 keV count rates lower than that expected at
confidence level $\gs 95\%(\gs99.7\%)$.  Large absorbing columns
densities are likely responsible for the hard spectrum of these
sources.  A deficit of very hard sources at fluxes
$\ls1-2\times10^{-13}$ \cgs is also evident in figure 1.  This can be
due to both an astrophysical and a technical reason. The first is a
redshift effect: the observed softness ratio of sources with similar
intrinsic absorbing column density increases with the redshift, as the
observed cut-off energy moves toward lower energies. The second is the
MECS reduced sensitivity to hard sources, due to the rapid increase of
the vignetting of the telescopes with the energy and with the off-axis
angle.  To quantify the latter effect we have computed loci of equal
4.5-10 keV count rate for a given flux, shown by dashed lines in
figure 1. The strong curvature of these lines toward low values of
(S-H)/(S+H) indicate that a large part of the deficit of faint hard
sources is probably due to this effect.  The curves are bent toward
high flux values at high (S-H)/(S+H) too, because the MECS sensitivity
is reduced for very soft sources by the berillium window, which
absorbs most photons below $\sim2$ keV.  The 4.5-10 keV sensitivity is
maximum for an unabsorbed power law spectrum of $\alpha_E=0.6-0.8$.

\begin{figure}[ht]
\centerline{
\hbox{
\psfig{file=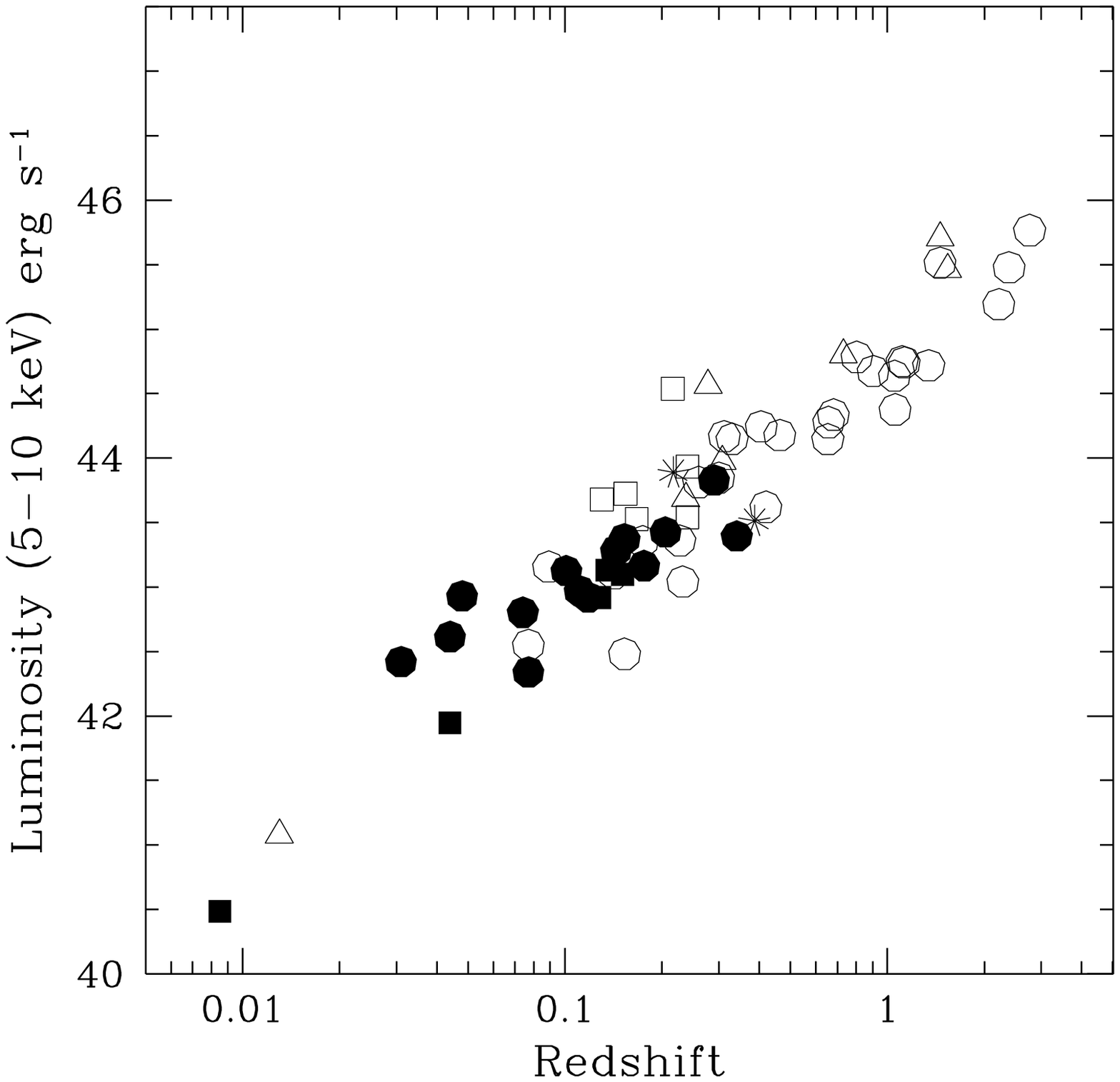, width=6cm, height=6cm}
\psfig{file=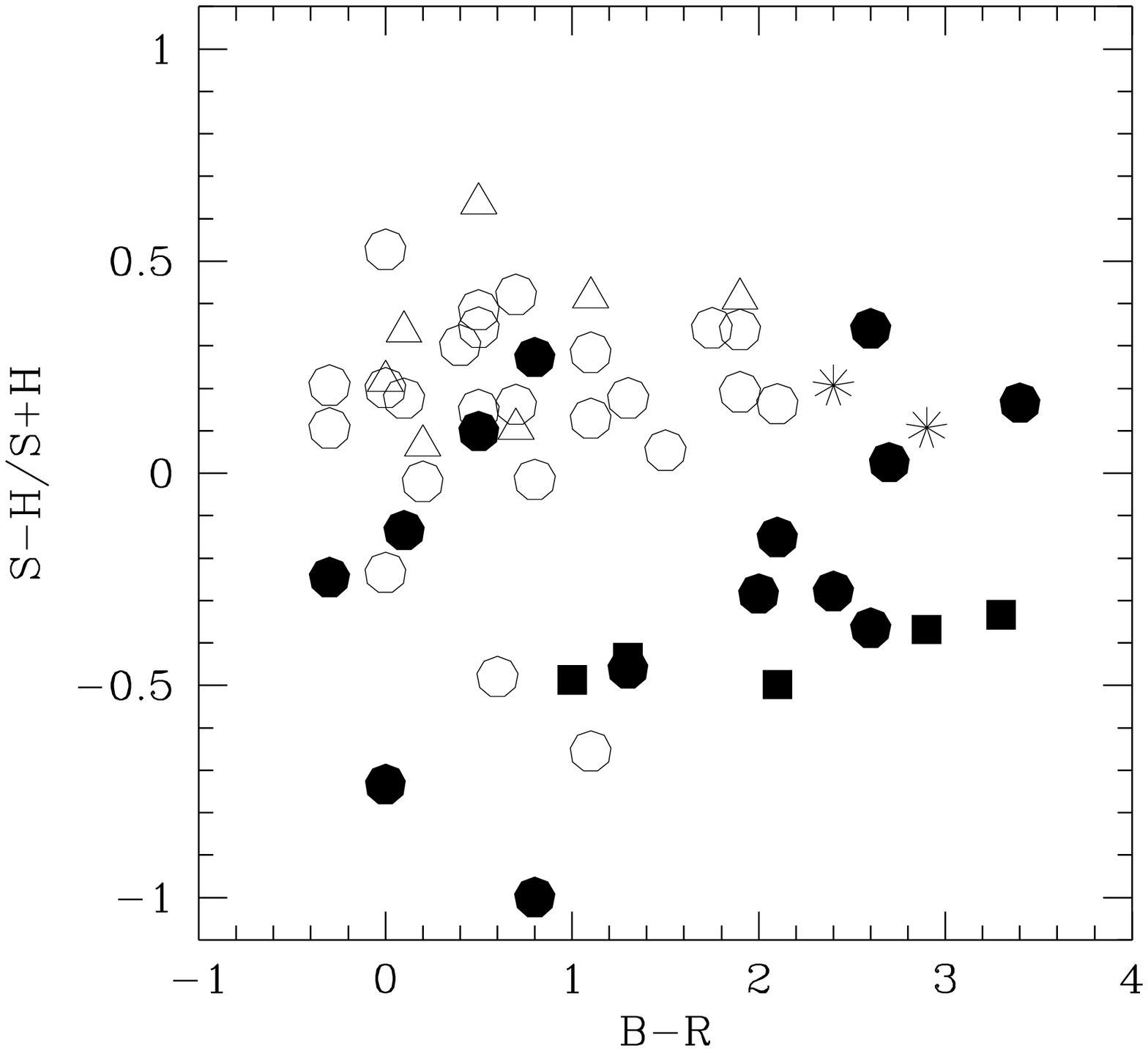, width=5.7cm, height=5.7cm }
}
}
\caption{a) The luminosity of the HELLAS sources as a function
of their redshift. Different symbols mark different sources:
open circles = broad line, `blue' continuum quasars and Sy1; stars=
broad line `red' continuum quasars; filled circles= type 1.8-1.9-2.0
AGN; filled squares= starburst galaxies and LINERS; open triangles=
radio-loud AGN; open squares= clusters of galaxies.
A cosmology with $H_0=70$ and $q_0=0.5$ has been used.
b) The (S-H)/(S+H) softness ratio versus the B--R color
}
\label{zlx}
\end{figure}

\section{Optical identifications}

Correlations of the HELLAS source catalog with catalogs of cosmic
sources provide 26 coincidences (7 radio-loud AGN, 13 radio-quiet
AGN, 6 clusters of galaxies), suggesting that most of the HELLAS
sources are AGN. Optical spectroscopic follow-ups have been performed on
about 45 HELLAS error-boxes, providing 36 new identifications (Fiore
et al. 1999, La Franca et al. 2000 in preparation).
The radio-quiet AGN sample includes:

%
i) 28  broad line blue continuum quasars 

ii) 2 broad line `red' continuum quasars

iii) 14 type 1.8-1.9-2 AGN

iv) 5 sources have optical spectra typical of LINERS of starburst
galaxies (all have strong [OII] emission).


\begin{table}
\label{lines}
\caption{\bf Mean (S-H)/(S+H) and B--R of HELLAS radio-quiet AGN}
\begin{tabular}{lcccc}
\hline
               & $<B-R>$ & $\sigma_{B-R}$ & $<(S-H)/(S+H)>$ & 
$\sigma_{(S-H)/(S+H)}$ \\
\hline
total AGN (49) & 1.2$\pm$0.2 & 1.0 & -0.03$\pm$0.05 & 0.34 \\  
broad line AGN (28) & 0.8$\pm$0.1  & 0.7 & 0.11$\pm$0.05 & 0.26 \\  
narrow line AGN (19) &1.7$\pm$0.3  & 1.1 &-0.26$\pm$0.08 & 0.34 \\  
\hline

\end{tabular}
 
\end{table}

\begin{figure}[ht]
\centerline{
\psfig{file=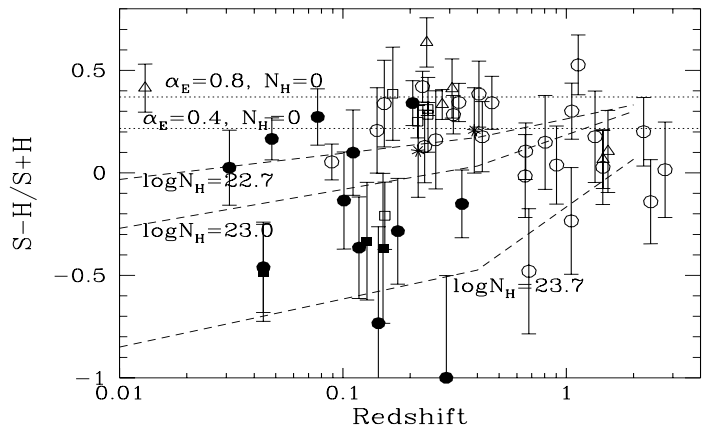, width=7.cm, height=11cm, angle=-90}
}
\caption{(S-H)/(S+H) versus the redshift for
the identified sources. Symbols as in figure 2. Dotted lines
show the expected softness ratio for a power law model with
$\alpha_E$=0.4 (lower line) and $\alpha_E=0.8$ (upper line).  Dashed
lines show the expectations of absorbed power law models (with
$\alpha_E=0.8$ and log$N_H$=23.7, 23.0, 22.7, from bottom to top) with
the absorber at the source redshift.  }
\label{hrtz}
\end{figure}

\begin{figure}[t]
\centerline{
\hbox{
\psfig{file=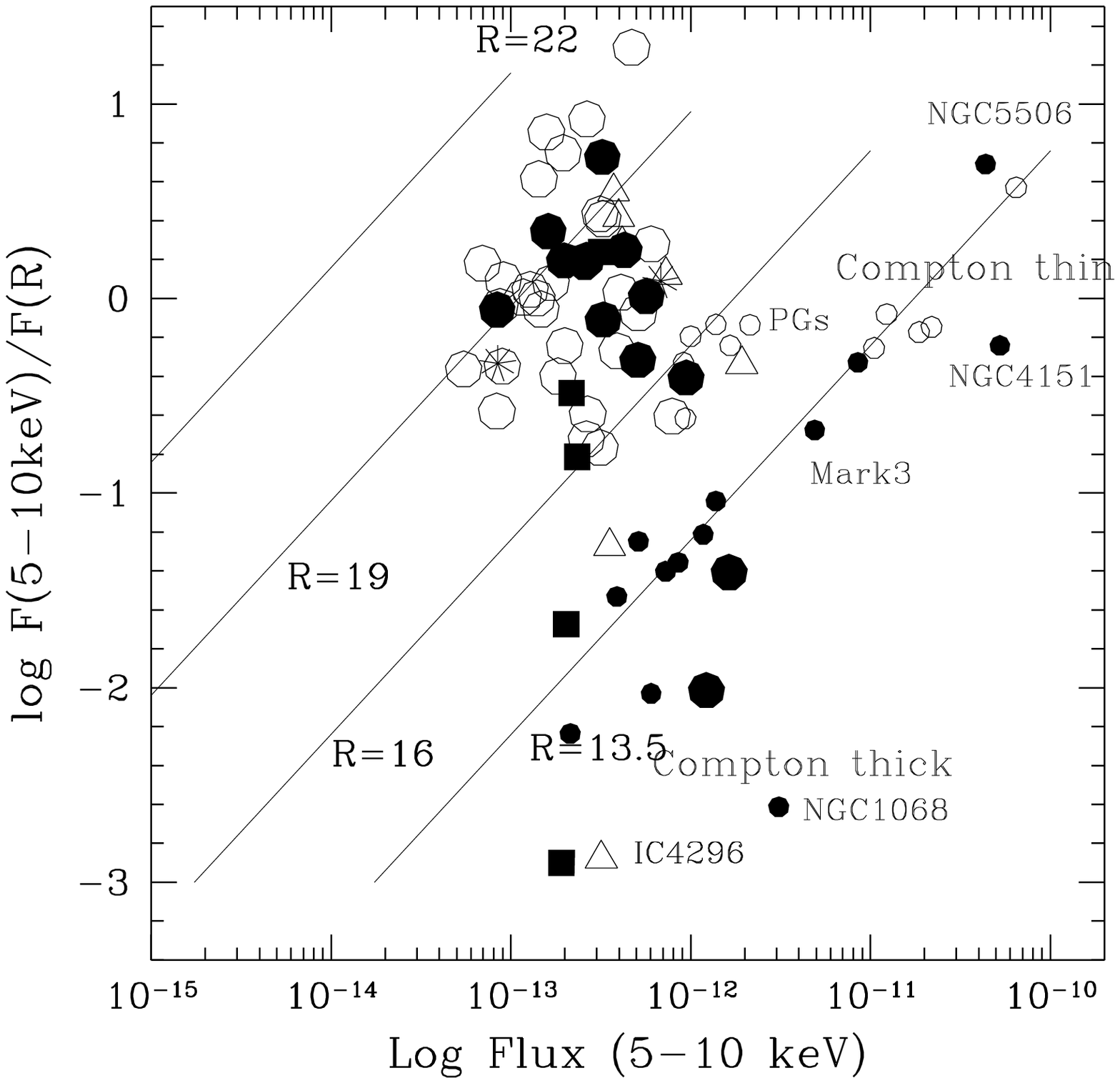, width=6.2cm, height=6.2cm}
\psfig{file=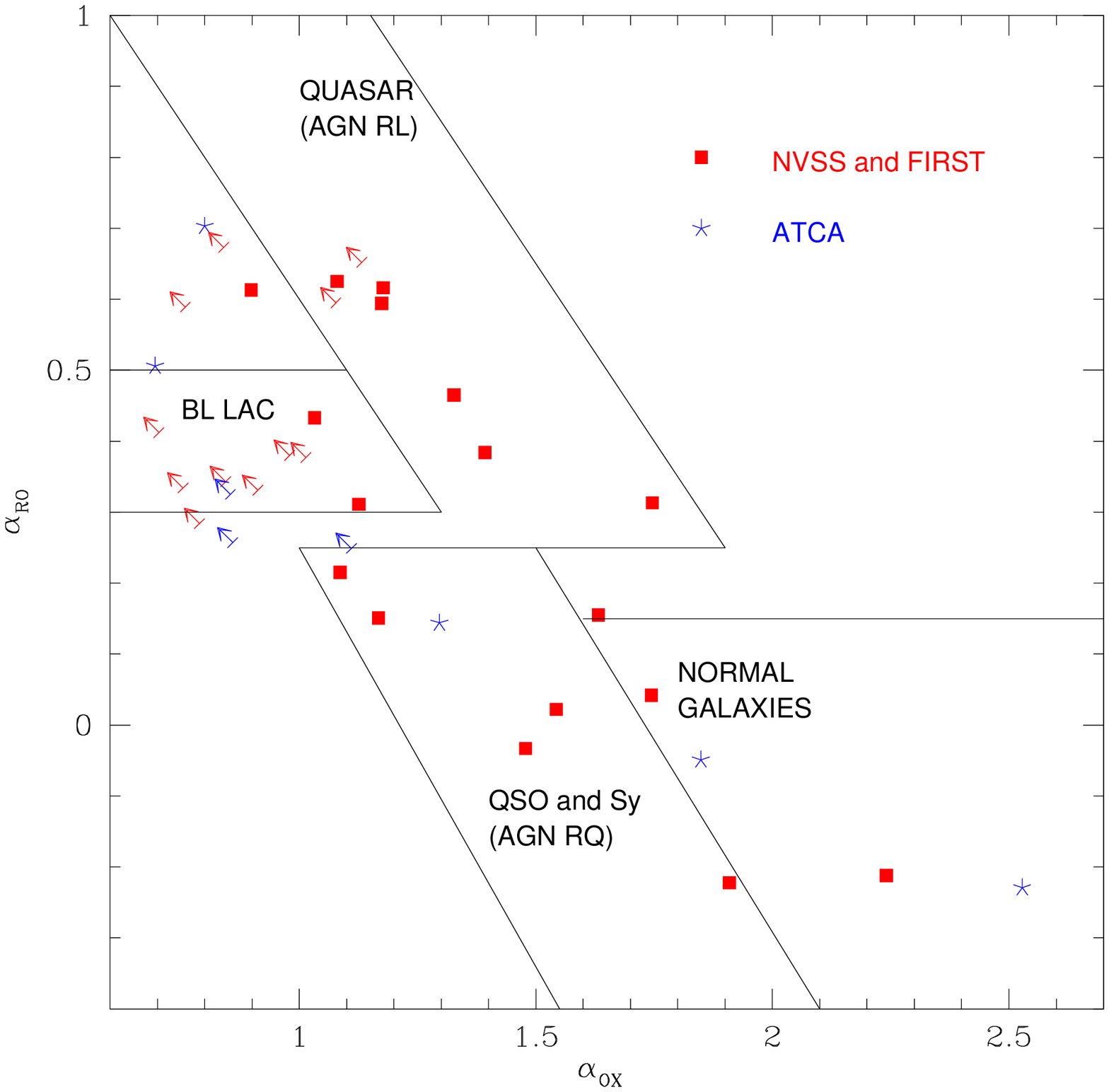, width=5.8cm, height=5.8cm}
}
}
\caption{a) X-ray to optical ratio versus the 5-10 keV flux.  
HELLAS AGN: big symbols; known AGN: small
symbols (symbols as in figure 2).  In Compton thick AGN
(log$N_H>24.3$), the nuclear emission is completely blocked in the 5-10
keV band. Thick lines identify regions of constant apparent R
magnitude. b) The $\alpha_{ro}$ versus the
$\alpha_{ox}$ for the HELLAS sources with a radio counterpart.
}
\label{radio}
\end{figure}

$\bullet$ The number of chance coincidences in 45 error boxes is
$<0.8-1.5$ for broad line quasar and $<4$ for narrow line AGN
(including in this category strong [OII] emission line galaxies like
LINERS and starburst galaxies).

$\bullet$ At least 5 of the narrow emission line AGN lie in small
groups and/or in interacting couples.

$\bullet$ About 1/3 of the error-boxes studied in detail contains no
`reasonable' (in terms of X-ray to optical ratio of known classes of
sources) counterpart to the X-ray source down to R=20.5

$\bullet$ Optical spectroscopy indicates a wide variety of spectra,
with a large fraction of ``intermediate'' objects (type 1.8-1.9 AGN,
`red' quasars).

Figure 2a) plots the luminosity of the identified HELLAS sources as a
function of their redshift. We have identifications of broad line
quasars up to z=2.76 and luminosity of $\sim10^{46}$ erg s$^{-1}$, and
of narrow line AGN up to z=0.4 and luminosity of $\sim10^{44}$ erg
s$^{-1}$. Figure 2b) plots the (S-H)/(S+H) of the identified sources
as a function of their B~--~R color. Note as the broad line AGN are
relatively well separated from the narrow line AGN in the diagram,
although the scatter in both B~--~R and (S-H)/(S+H) is large for both
class of AGN (also see Table 1).  Both (S-H)/(S+H) and B~--~R of broad
line AGN are different (at the $\sim3\sigma$ level) from that of
narrow line AGN, which, on average, have a lower (S-H)/(S+H), and are
therefore likely to be more X-ray absorbed, and an higher B~--~R, and
are therefore subjected to a greater extinction.

(S-H)/(S+H) is plotted as a function of the redshift in figure 3 for
the 53 identified sources detected far from the berillium strongback
supporting the MECS window.  The dotted lines represent the
expectation of unabsorbed power law with $\alpha_E=0.4$ and 0.8.  The
dashed lines represent the expectations of a power law absorbed by
columns of $5\times10^{22}$, $10^{23}$ and $5\times10^{23}$ cm$^{-2}$
respectively, in the source frame.  Note that the softness ratios of
constant column density models strongly increases with the redshift.
Most of the narrow line AGN have (S-H)/(S+H) inconsistent with that
expected from a power law model with $\alpha_E=0.4$.  Absorbing
columns, of the order of $10^{22.5-23.5}$ cm$^{-2}$, are most likely
implied. Note also that some of the broad line AGN have (S-H)/(S+H)
inconsistent with that expected for a $\alpha_E=0.8$ power law, in
particular at high redshift. The (S-H)/(S+H) of the 24 broad line AGNs
is marginally anticorrelated with z (Spearman rank correlation
coefficent of -0.364 for 22 degrees of freedom, corresponding to a
probability of 92\%). The number of sources is not large enough to
reach a definite conclusion, but it is interesting to note that this
correlation goes in the opposite direction than expected. In fact, the
ratio of the optical depth in the optical band, due to dust
extinction, to that in the X-ray band, due to photoelectric
absorption, should scale as $(1+z)^4$. Highly X-ray obscured broad
line blue continuum quasar can exist only if their dust to gas ratio
or their dust composition strongly differs from the Galactic one (also
see Maiolino, this meeting). Similar results have been recently found
in ASCA samples by Akiyama et al. (1999) and Della Ceca et al. (this
meeting). XMM-Newton and Chandra follow-up observation may easily
confirm or disregard a significant absorbing column in these high z
broad line quasars.

\section{Spectral Energy Distribution}

\subsection{X-ray to optical color}

Figure 4 shows the hard X-ray (5-10 keV) to optical (R band) flux
ratio as a function of the X-ray flux for the identified HELLAS
sources and a sample of relatively bright, nearby AGN observed by
BeppoSAX (Seyfert 1 galaxies, Seyfert 2 galaxies, PG quasars with
z$<$0.4).  The X-ray to optical ratio of the HELLAS sources is similar
to that of the X-ray brightest objects in the local universe (with the
exception of X-ray selected blazars, like the HBL, which have higher
X-ray to optical ratio, but also a relatively strong radio emission).
While supporting the robustness of our identifications, this suggests
that roughly one third of the hard X-ray background is due to sources
similar to local Seyferts and quasars.

\subsection{Near infrared to optical colors}

Photometric infrared and optical observations of 10 HELLAS sources
have been carried out using the Telescopio Nazionale Galileo (TNG;
Maiolino et al. 2000).  The sample includes 4 broad line `blue'
quasars, 2 broad line `red' quasars, 3 type 1.9 AGN and 1 LINER.  The
B, R and J photometry of the 2 `red' quasars and the 4 narrow line
galaxies is dominated by the emission from the host galaxy. AGN
contribution is observed in the K band, especially in the 2
`red' quasar. This means that many, if not most of the objects making
the hard X-ray background cannot be distinguished from normal galaxies
using optical and near-IR photometry. In fact only the 4 broad line
blue quasar would have passed the color criterion of the PG catalog,
thus indicating that a large fraction of the accretion power in the
Universe may actually have been missed in optical color survey such as
the BQS (U~--~B$<-0.44$).  Multicolor photometry techniques based on
galaxy templates (e.g. Giallongo et al.  1998) may be efficiently used
to assess the redshift of the hard X-ray selected sources. This will
be more and more important when large samples of faint X-ray sources
will be available from Chandra and XMM observations, and optical
spectroscopic identification of all of them will not be feasible.

\subsection{Radio to optical to X-ray broad band spectral indices}

Nearly all the 147 HELLAS sources have been observed in either the
NVSS and FIRST surveys or by our collaboration using ATCA.  In
particular, we observed with ATCA 20 sources obtaining 8 5$\sigma$
detections (40 \% of the sample) at a 5 GHz flux limit of 0.5
mJy. This fraction is higher than in Radio follow-ups of Einstein and
ROSAT X-ray surveys (Ciliegi et al. 2000 in preparation).  This is
mainly due to the lower radio-to-X-ray flux limit ratio in our survey.
A similar fraction of radio detections was found by Akiyama et
al. (2000) in their correlation of the ASCA Large Sky Survey sources
with the FIRST catalog, which have a radio-to-X-ray flux limit ratio
similar to ours. Figure 4b) shows the radio-to-optical broad band
spectral index as a function of the X-ray-to-optical index (arrows
identify sources without a R$<$20 counterpart in the small Radio error
box).  Note that many of the HELLAS detections have $\alpha_{RO}$ and
$\alpha_{XO}$ consistent with that of radio quiet AGN.

\section {Conclusions}

A large area, hard X-ray survey performed with BeppoSAX has found a
population of faint obscured AGN. The X-ray data, complemented by
optical, near-IR and radio follow-ups, indicate that the majority of
these sources are ``intermediate'' AGN i.e. type 1.8-1.9 AGN, `red'
quasars, obscured in X-ray by columns of the order of $10^{22.5-23.5}$
cm$^{-2}$, but showing a wide dispersion in optical extinction.  The
sample of identified HELLAS AGN contains higher redshift analogs of
nearby Seyfert galaxies and quasars. Furthermore, we find marginal
evidence for a population of X-ray obscured quasars at z$\gs0.5$
showing broad lines in their optical spectra.

\begin{acknowledgements}
We thank the BeppoSAX SDC, SOC and OCC teams for the successful
operation of the satellite and preliminary data reduction and
screening. This research has been partly supported by ASI 
ARS/99/75 contract and MURST Cofin-98-032 contract.
\end{acknowledgements}

\end{document}